\documentstyle[epsfig]{aipproc}

\def\pdev(#1,#2;#3){\left(
           \kern-0.1em{\partial #1\over\partial #2}
           \kern-0.1em\right)_{\kern-.2em #3}}
\def\batio3{BaTiO$_3$}
\def\ave(#1){\langle#1\rangle}
\def\corr(#1,#2){\ave{\Delta{#1}\Delta{#2}}}

\begin{document}
\title{Temperature-dependent dielectric response of BaTiO$_3$ from
first principles}

\author{Alberto Garc\'{\i}a$^*$ and David Vanderbilt$^{\dagger}$}
\address{$^*$Departamento de F\'{\i}sica Aplicada II, Universidad del Pa\'{\i}s
Vasco\\ Apdo. 644, 48080 Bilbao, Spain, email: wdpgaara@lg.ehu.es\\
$^{\dagger}$Department of Physics and Astronomy, Rutgers University\\
136 Frelinghuysen Road, Piscataway, NJ 08854-8019, USA, email: dhv@physics.rutgers.edu}

\maketitle

\begin{abstract}
Monte Carlo simulations with an effective Hamiltonian parametrized
from first principles are performed to study the dielectric response
of \batio3\ as a function of temperature, with particular emphasis on
the behavior of the dielectric constant near the transition from the
ferroelectric tetragonal phase to the paraelectric cubic phase.
\end{abstract}

The peculiar dielectric properties of ferroelectric materials stem
from the coupling of the electric field to polar distortions of the
crystal lattice. In one common scenario, the progressive softening of
a lattice vibrational mode in the neighborhood of a phase transition
brings about a dramatic rise in the value of the dielectric
constant. The resulting very large relative permitivities (in the
thousands range) have found important technological applications. From
a practical point of view, however, the development of materials with
a desired response faces the difficulty of trying to separate
experimentally the influences of many effects: composition, structure,
domain configuration, etc. There are also open theoretical issues
such as whether the relevant phase transitions are indeed associated
to the above-mentioned ``soft-mode'' (and are thus of ``displacive''
character) or exhibit ``order-disorder'' characteristics. The use of
first-principles calculations can help in both fronts. They allow
the study of the response of a system under ``controlled'' conditions
that would be very difficult or impossible to
realize in the laboratory. Besides, they provide a microscopic view of
the materials which is simply not available experimentally and, unlike
simplified models, they are tailored to the detailed chemical
composition of the system.

In the last few years much progress has been made in the
computational study of ferroelectric materials through the use of
effective Hamiltonians which contain the physically relevant degrees
of freedom of the structure. The effective Hamiltonians are
constructed on the basis of first-principles calculations, and the
statistical mechanics of the system is then studied by Monte Carlo
simulation. Calculations of the phase transition
sequence~\cite{agarcia:mc-batio3} and ferroelectric domain
walls~\cite{agarcia:wall} in \batio3, and of the ferroelectric
transition in PbTiO$_3$~\cite{agarcia:mc-pbtio3}, have proved the
usefulness of this approach.

In this work we present the first calculations of the
temperature-dependent dielectric response of the perovskite \batio3
from first principles, with particular emphasis on the behavior of the
dielectric tensor in the vicinity of the phase transition from the
tetragonal ferroelectric to the cubic paraelectric phase.

The basic ingredient of the Metropolis Monte Carlo algorithm is the
generation of a sequence of states which are distributed according to
the Boltzmann probability ${\rm Prob}_j=\exp(-\beta U^j)$, where $U^j$ is
the energy of the state $j$ and $\beta=(kT)^{-1}$. A state $j_{n+1}$
is added to the sequence after state $j_n$ on the basis of a
transition probability $\pi(n,n+1)={\rm min}\{1,\exp[\beta
(U^{j_{n}}-U^{j_{n+1}})]\}$. In our method the energy $U$ of the
system is represented by means of an effective Hamiltonian $H_{\rm
eff}$ which is basically a Taylor expansion of the energy surface
around the high-symmetry cubic perovskite structure.  $H_{\rm eff}$ is
written in terms of the dynamical variables which are relevant to the
low-energy distortions: the amplitudes \{{\bf u}\} of the local modes
(three degrees of freedom per unit cell) which represent the ``soft''
transverse optical phonon and are directly related to the polarization
{\bf P} of the crystal~\cite{agarcia:anharm} [${\bf P}=(Z^*/V)\sum{\bf
u}$, where $Z^*$ is the mode effective charge and $V$ is the cell
volume]; a set \{{\bf v}\} of displacement variables representing the
acoustic modes; and the six components of the homogeneous strain
$\eta$. The parameters of the energy expansion, including those for
the on-site
local-mode self-energy, the interaction between local modes (both
short-range and dipole-dipole), the elastic energy, and the local
mode-elastic coupling, are computed using highly accurate
first-principles LDA calculations with Vanderbilt ultrasoft
pseudopotentials~\cite{agarcia:uspp}.  More details about the
construction of the effective Hamiltonian can be found in
Ref.~\onlinecite{agarcia:mc-batio3}. The extension of the standard
Metropolis Monte Carlo algorithm to include the effects of stress
$\sigma$ and electric field {\bf E} involves replacing the Boltzmann
probability factor $\exp(-\beta U^j)$ by
$\exp[-\beta(U^j-V_0\sigma_\nu\eta^j_\nu-E_i{\cal P}^j_i)]$ in the
acceptance criterion for state $j$ (here ${\cal P}_i=VP_i$ is the
$i$th component of the net dipole moment of the crystal and $V_0$ is
the volume for zero strain). For a given
temperature, stress, and field, the strain $\eta$ and the mode
variables are allowed to fluctuate, their average values determining
the strain and net polarization of the system. This extended framework
has recently been used to study the piezoelectric response of \batio3\
as a function of temperature, and to illustrate the influence of
electric fields on the phase diagram of this
material~\cite{agarcia:piezo}.

Here, we use this approach to compute the
dielectric response of the cubic and the tetragonal (ferroelectric)
phases of \batio3. The tetragonal phase is stable from approximately 278K to
403K and exhibits a spontaneous polarization that we take to be along
the $z$ axis. The linear dielectric response coefficients (dielectric
tensor) are given, in S.I. units, by
$\varepsilon_{ij}=\varepsilon_0(1+\chi_{ij})$, where $\chi_{ij}$ is
the dimensionless dielectric susceptibility defined by
\begin{equation}
\chi_{ij}
={1\over\varepsilon_0}\pdev(P_i,E_j;\sigma,T)
\simeq {1\over V\varepsilon_0}\pdev({\cal P}_i,E_j;\sigma,T)
={1\over V\varepsilon_0}\beta\,
    (\ave({\cal P}_i{\cal P}_j)-\ave({\cal P}_i)\ave({\cal P}_j))\, .
\label{eq:agarcia:chi}
\end{equation}
The approximate equality reflects the neglect of the field derivative
of the volume~\cite{agarcia:dVdE}. Here the averages are to be
computed using the extended Boltzmann factor defined
above~\cite{agarcia:explain-corr}.  According to these equations, one
could either compute the linear dielectric response from direct calculations
of the average polarization as a function of electric field (``direct
approach''), or from an analysis of the statistical correlation
between polarization components (``correlation approach'').

\begin{figure} % 
\centerline{\epsfig{file=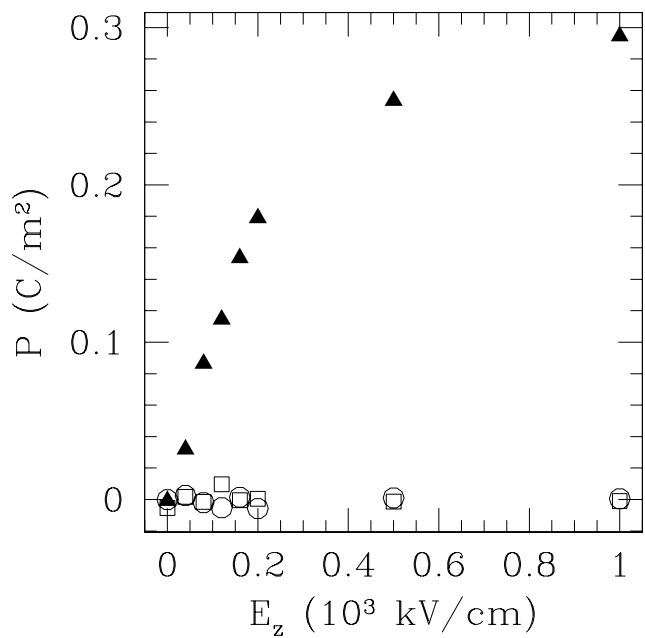,height=3.5in,width=3.5in}}
\vspace{10pt}
\caption{Average polarization vs.\ electric field for the cubic phase
of \batio3\ at a rescaled temperature of 500K. Solid triangles and
open circles and squares represent the $z$, $x$, and $y$ components of
the polarization, respectively.}
\label{fig:agarcia:p_vs_e}
\end{figure}

As an example of the direct approach, we show in
Fig.~\ref{fig:agarcia:p_vs_e} the results of a series of simulations
for the cubic phase in which an electric field of progressively
greater magnitude is applied along the $z$ direction.  For each field
value, the simulation box (a cube with $10\times 10\times 10=1000$
unit cells) was allowed to equilibrate for 2$\times10^4$ Monte Carlo sweeps
(MCS)~\cite{agarcia:mcs} and polarization averages were taken over
another 2$\times10^4$ MCS.  A fit to the $P_z$ vs.\ $E_z$ curve in the linear
region corresponding to field strengths up to approximately 150 kV/cm
can be used to extract the dielectric susceptibility.  For higher
fields, nonlinear effects are clearly present. It is important to
realize that the nonlinearity is not put in explicitly, as the only
extra term in the simulation is $-E_i\Delta {\cal P}_i$, which is linear in
the field. The nonlinearity appears through the terms of higher order
in the local mode variables $\{{\rm\bf u}\}$ (and their coupling to
the strain) in the effective Hamiltonian. A closer analysis of these
effects could form the starting point of an investigation of the nonlinear
dielectric response in this material.

\begin{figure} % 
\centerline{\epsfig{file=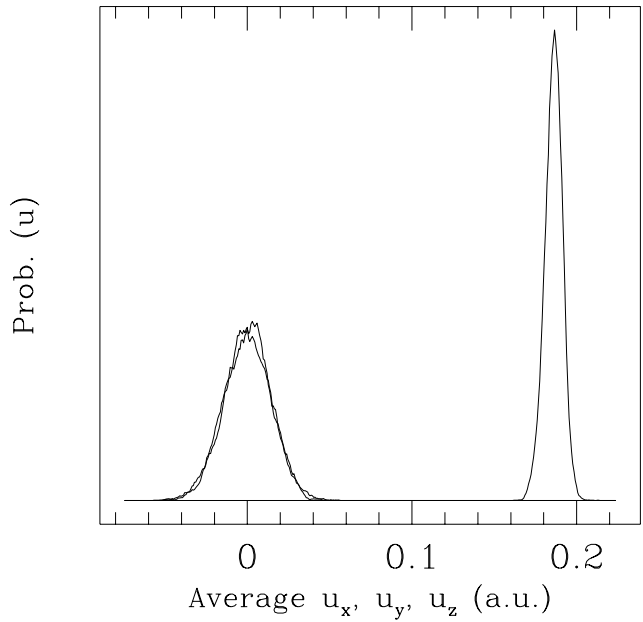,height=3.5in,width=3.5in}}
\vspace{10pt}
\caption{Probability histograms for the local mode components in the
tetragonal phase, obtained from a run with 3$\times10^5$ MCS and an $L=14$
simulation box (rescaled temperature: 355 K). The vertical scale is
arbitrary.}
\label{fig:agarcia:histo}
\end{figure}

The computation of linear response coefficients from the correlation
approach requires only one simulation at zero field,
although relatively long runs (of at least $10^5$ MCS) are
needed to obtain good statistics. Also, the quality of the calculated
correlations should improve with the size of the simulation box
(recall that the relations in Eq.~\ref{eq:agarcia:chi} are strictly
valid only in the thermodynamic limit). We have therefore performed
our calculations using larger boxes, with $12\times 12\times 12=1728$
($L=12$) and $14\times 14\times 14=2744$ ($L=14$) unit
cells. Figure~\ref{fig:agarcia:histo} shows histograms for the
averages of the three components of the local mode in the tetragonal
phase, computed with an $L=14$ box and 3$\times10^5$ MCS. The profiles are
quite clean and gaussian-looking. The diagonal components of $\chi$
(which are the relevant ones for the tetragonal and cubic phases) are
related to the width of the statistical distribution of the
corresponding component of ${\bf P}$, and thus to the standard
deviation of the system-average of the local mode amplitudes. Wider
distributions (such as those for the $x$ and $y$ components in
Fig.~\ref{fig:agarcia:histo}) indicate larger values for the
corresponding components of the dielectric tensor.

Note that the correlation method provides
information about the whole tensor from a single run, which is
particularly economical when dealing with lower-symmetry phases. In
these cases the use of the direct method becomes more cumbersome. For
example, the determination of $\chi_{33}$ and $\chi_{11}$ for the
tetragonal phase would involve two separate series of direct
calculations, for varying $E_z$ and $E_x$, respectively. In the latter,
extra care would be needed at temperatures close to the transition to
ensure that a transverse field does not cause a switch of the
spontaneous polarization from the $z$ to the $x$ direction.

\begin{figure} %
\centerline{\epsfig{file=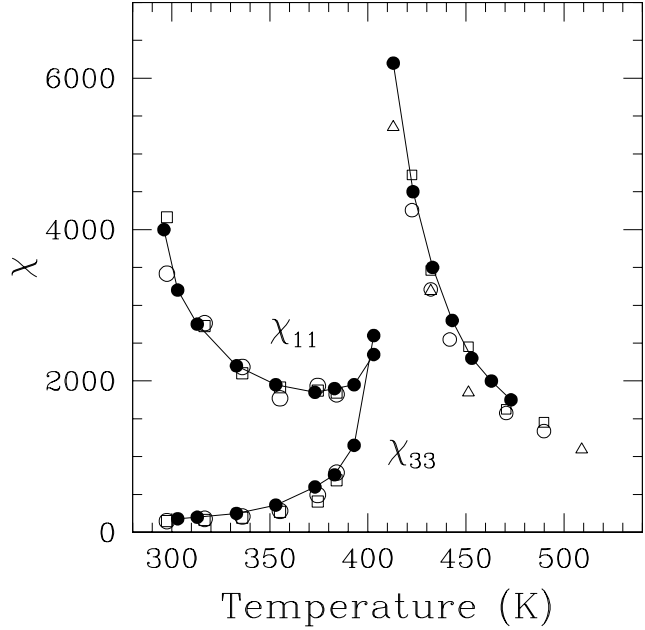,height=3.5in,width=3.5in}}
\vspace{10pt}
\caption{Constant-stress dielectric susceptibility for the tetragonal
and cubic phases of \batio3. Solid circles are experimental data from
Ref.~\protect\onlinecite{agarcia:li-exp}. Open symbols represent computed
values, plotted vs.\ rescaled temperature (see text).
Triangles are from direct P vs. E simulations, and circles and
squares are from correlation analysis with $L=12$ and $L=14$ boxes,
respectively.}
\label{fig:agarcia:eps}
\end{figure}

Before presenting our results we must discuss an important point
regarding the temperature scale.  An effective Hamiltonian based on a
{\sl finite} Taylor expansion of the energy should not be expected to
reproduce perfectly the behavior of the material at relatively high
temperatures. In particular (see Ref.~\onlinecite{agarcia:mc-batio3})
the theoretical transition temperatures are progressively shifted
downwards with respect to the true ones~\cite{agarcia:tcs}, the
agreement worsening as the temperature increases. This shift might be
related to the neglect of higher order terms in the interaction
between local modes in $H_{\rm eff}$, as transition temperatures
depend basically on the details of the interaction (as a simple
example, recall the Ising model, for which $T_c$ is proportional to
the spin-spin coupling $J$)~\cite{agarcia:lda_error}. In order to
provide a better comparison of our results to experiment, we have
therefore linearly rescaled the theoretical temperatures so that the
end points of the range of stability of the tetragonal phase coincide
with the experimental $T_c$'s.  The rescaling is also extrapolated
into the range of stability of the
cubic phase. By fixing the points at which phase transitions occur, we
are able to focus on the consequences of lattice instability for the
dielectric response.  The more
important, low-energy regions of the energy surface, which are
correctly parametrized by our $H_{\rm eff}$, presumably play
the most important role in determining this response.

Figure~\ref{fig:agarcia:eps} shows the computed constant-stress
dielectric susceptibility for the cubic and tetragonal phases,
together with recent experimental data~\cite{agarcia:li-exp} for the
low-frequency dielectric response of
\batio3~\cite{agarcia:hi_freq}. The agreement is excellent, with the
simulations reproducing in detail all the features of the observed
behavior, including the large anisotropy of the dielectric tensor in
the tetragonal phase. Our data from the direct simulation of the {\bf
P} vs.\ {\bf E} curves (cubic phase only) and statistical correlation
with various box sizes are comparable within the aforementioned
limitations of a finite simulation box.
\begin{figure} %
\centerline{\epsfig{file=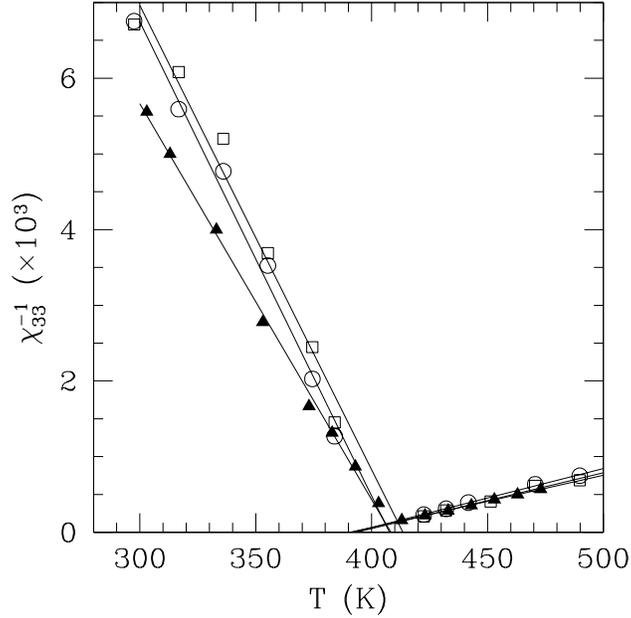,height=3.5in,width=3.5in}}
\vspace{10pt}
\caption{Fit of the inverse susceptibility to a Curie-Weiss
form. Solid triangles are experimental data from
Ref.~\protect\onlinecite{agarcia:li-exp}. Open circles and squares represent
values computed from correlation analysis with $L=12$ and $L=14$
boxes, respectively.}
\label{fig:agarcia:Curie}
\end{figure}
Near the transition temperature $T_c$ the pseudo-divergent behavior of
$\chi_{33}$ can be
approximated to a Curie-Weiss form $\chi_{33}^{-1}=C(T-T_0)$, where $T_0$ is
a temperature close but not identical to $T_c$
(Figure~\ref{fig:agarcia:Curie}). The fitted values of the Curie
temperature $T_0$ extracted from the tetragonal and cubic phases are very
similar, reflecting the fact that the transition is only weakly
first-order. The values of the Curie-Weiss constants above and below
the transition are not related in a simple way in a first-order
transition, but are predicted from mean-field arguments (Landau
theory~\cite{agarcia:landau}) to satisfy $C_{below}=-2C_{above}$ for a
second-order transition.  Simulations with a three-dimensional
$\phi^4$ model~\cite{agarcia:radescu} (which could be seen as a
simplified form of our effective Hamiltonian) have shown that this
mean-field relationship between the $C$ constants is only valid in the
limit of a pure displacive phase transition, and that
$|C_{below}|/2C_{above}$ increases with the degree of order-disorder
character. Our data (and experiment) show that $|C_{below}|/2C_{above}
\simeq 4$.  Assuming the deviations from mean-field behavior are similar
for our $H_{\rm eff}$ and the $\phi^4$ model, this result
would be consistent with the analysis
in Ref.~\onlinecite{agarcia:mc-batio3} pointing to a relatively strong
order-disorder character of the cubic-tetragonal transition in
\batio3.

In conclusion, we have shown how the temperature-dependent dielectric
response of a system can be computed from first principles using an
effective Hamiltonian and Monte Carlo simulations. As an application,
we have presented the first calculations of the dielectric response of
\batio3\ as a function of temperature. The analysis of the behavior of
the dielectric susceptibility in the vicinity of the cubic-tetragonal
transformation gives evidence for a certain degree of order-disorder
character in the transition.

This work was supported in part by the UPV research grant
060.310-EA149/95 and by the ONR Grant N00014-97-1-0048. We
thank J.M. Perez-Mato and Karin Rabe for useful comments.


\begin{references}
\bibitem{agarcia:anharm} To relate the net macroscopic
polarization to the soft-mode average, thus excluding the
contributions from other polar modes, is equivalent to assuming that
there are no anharmonic effects in the dynamics of the other modes, so
that their average amplitude is zero.
\bibitem{agarcia:dVdE} The linear change in volume is strictly zero by
symmetry in the cubic phase, and in the tetragonal phase for fields
perpendicular to the ferroelectric axis.  If an electric field is
applied along the ferroelectric axis, the longitudinal elongation and the
transverse contraction are such that the net change in volume is very
nearly zero (the piezoelectric coefficients satisfy $d_{33}\simeq
-2d_{31}$).
\bibitem{agarcia:explain-corr} The relationship between $\chi$ and the
polarization correlations can be simply obtained
by differentiation of the expression $\Lambda\ave(X)=\sum_j{X^j
{\rm Prob}_j}$, where Prob$_j$ is the extended Boltzmann factor,
$\Lambda$ is the extended partition function, and $X$ is any
component of the net polarization.
\bibitem{agarcia:mcs} A Monte Carlo sweep is completed after each local
variable is considered for a ``flip attempt'' and each component of
the homogeneous strain suffers $2L+1$ attempted changes, where $L$ is
the linear size of the simulation box.
\bibitem{agarcia:tcs} For the rhombohedral to orthorhombic to tetragonal
to cubic transition sequence, the experimental $T_c$'s are
respectively 187K, 278K, and 403K, while the theoretical ones
are 197K, 230K, and 295K.
\bibitem{agarcia:lda_error} Another possible source of the error in
the computed values of transition temperatures is the underestimation
by the LDA of the equilibrium lattice constant. The simulations are
run at a negative effective pressure to compensate for this effect (see
Ref.~\protect\cite{agarcia:mc-batio3}).
\bibitem{agarcia:hi_freq} At high frequencies the strain is not able to
follow the polarization and the experiment measures the {\sl
constant strain} dielectric susceptibility.
\end{references}
\end{document}